# Comparison of different algorithms for under-sampled image reconstruction
(Student paper)

Ana Šćekić, Dražen Jelić, Nemanja Ševaljević, Melvudin Hot
University of Montenegro, Faculty of Electrical Engineering, Podgorica
ana.scekic96@gmail.com, drazen.jelic@gmail.com, nemanjasevaljevic@gmail.com, dinohot@ymail.com

**Introduciton**

The Compressive Sensing (CS) [1]-[2] as a novel acquisition approach that finds its usage in image processing. The hypothesis like this one assures signal recovery with high quality from decreased number of samples compared with the number required by the Nyquist – Shannon sampling theorem. It includes a gathering of strategies for representing a signal that are based on the predetermined number of estimations and after that signal reconstruction. The CS has been broadly utilized and applied in numerous applications including computed tomography, WiFi communication, image processing and camera design. Complex mathematics is developed in order to ensure signal reconstruction from relatively small information [1]-[21]. Two commonly used groups of the algorithms are convex optimization and greedy approaches.

**Reconstruction Algorithms**

The CS reconstruction methods [3]-[14] provide signal reconstruction form a limited number of measurements or available signal samples. The problem is how to effectively restore the original signal from compressed data. For that purpose, the reconstruction algorithms are used. The reconstruction algorithms for sparse signal recovery in CS can be broadly divided into six types as show in Fig.1.

**Convex Optimization**

This class of algorithms solve the problem of convex optimization using linear programming to obtain reconstruction [3]. Those methods are computationally complex. Some examples of such algorithms are:

1) Basis Pursuit
2) Basis Pursuit De-Noising (BPDN)
3) Least Absolute Shrinkage and Selection Operator (LASSO)
4) Least Angle Regression (LARS)

Basis Pursuit is based on signal decomposition into an "optimal" superposition of dictionary elements. Under term "optimal" it means that the solution has the smallest $l_1$ norm of coefficients among all such decompositions. The Basis Pursuit is applied in problems such as total variation denoising, illposed problems, abstract harmonic analysis etc.

**Greedy Iterative Algorithm**

Greedy algorithms [4], [12], [13] are popular and widely used nowadays. This class of recovery algorithms are fast reconstruction approach and provides relatively not complex mathematical framework. This class of algorithms offers solution for reconstructing by finding the sparsest solution, step by step (in an iterative way). The most important greedy algorithms include:

1) Matching Pursuit
2) Orthogonal Matching Pursuits (OMP)

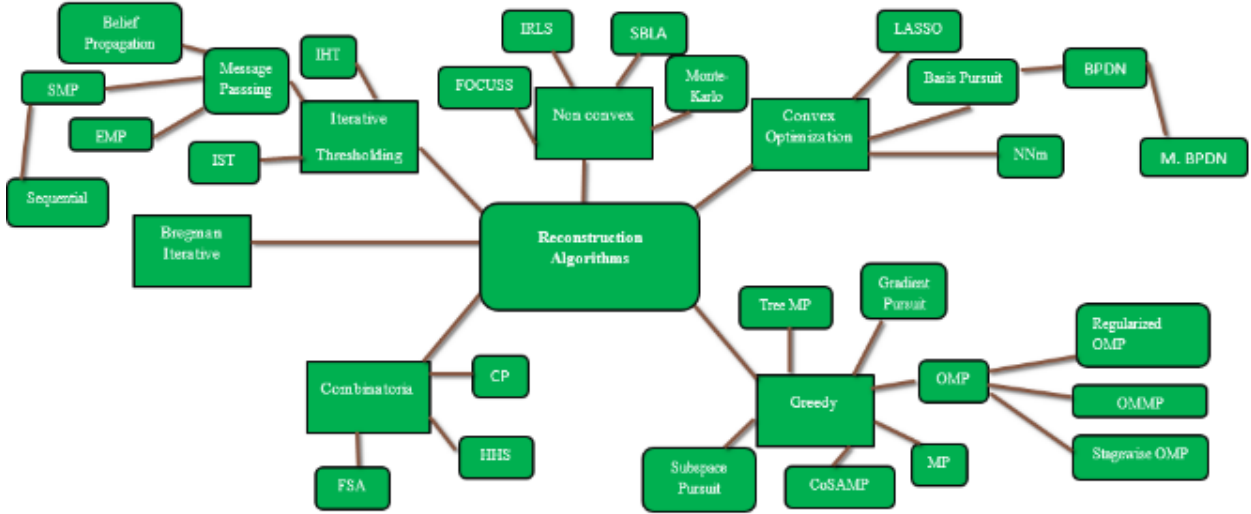

Figure 1. Reconstruction Algorithms

They are most used because of their low implementation cost and high speed of recovery. However, when the signal is not much perfectly sparse, recovery becomes costly. In such situations, improved versions (OMPs) are made as:

1) Regularized OMP
2) Stagevise OMP
3) Compressive Sampling Matching Pursuits (CoSaMP)
4) Subpace Trial
5) Gradient Pursuits and
6) Orthogonal Multiple Matching Pursuit

**OMP algorithm**

Orthogonal Matching Pursuit Algorithm (OMP) [10], [12] is a greedy compressed sensing recovery algorithm. It is based on iterative selection of the best fitting column of the sensing matrix. A least squares (LS) optimization is then performed in the subspace spanned by all previously picked columns. This method is less accurate than the convex optimization approaches, but its advantage is a lower computational complexity. Some a priori information about the signal has to be known, such as the sparsity $K$, the vector of measurements $\mathbf{y}$ and the sensing matrix $\mathbf{A}$.

OMP is a reconstruction algorithm that in each iteration searches for the maximum correlations between the measurements and the matrix $\mathbf{A}=\mathbf{\Phi\Psi}$. Matrix $\mathbf{A}$ is called dictionary, while $\mathbf{\Psi}$ is transform matrix and $\mathbf{\Phi}$ is measurement matrix. The columns of matrix $\mathbf{A}$ are called atoms. Usually matrix $\mathbf{A}$ is obtained as a random partial Fourier transform matrix. The algorithm is described in the following steps:

1: $\mathbf{r_0} \leftarrow \mathbf{y}$
2: $\Omega_0 \leftarrow \emptyset$.
3: $\Theta_0 \leftarrow []$
4: for i=1..,K
5: $\omega_i \leftarrow \arg\max |\{r_{i-1}, A_j\}|$
6: $\Omega_i \leftarrow \Omega_{i-1} \cup \omega_i$
7: $\Theta_i \leftarrow [\Theta_{i-1}\ A_{\omega i}]$
8: $\mathbf{x}_i = \arg\min_x \|\mathbf{r}_{i-1} - \mathbf{\Theta}_i\mathbf{x}\|_2^2$
9: $\mathbf{a}_i \leftarrow \mathbf{\Theta}_i\mathbf{x}_i$
10: $\mathbf{r}_i \leftarrow \mathbf{y} - \mathbf{a}_i$
end for
return $\mathbf{x}_k, \mathbf{a}_k, \mathbf{r}_k, \Omega_k$.

First step of the algorithm is setting the initial residual vector $\mathbf{r}_0$ to measurements vector $\mathbf{y}$, where $\mathbf{y}=\mathbf{\Phi f}$. Next step is to set initial indices $\Omega \leftarrow \emptyset$. Third step is to set matrix of chosen atoms by adding a column from $\mathbf{A}$. Afterwards the new approximation of data and the new residual are calculated by removing the contribution of already chosen atoms. The final signal has nonzero values that are equal to the components in $\mathbf{x}_k$.

**Total-Variation Method**

Total-variation method [10], [12], [16] can be used for denoising and restoring of noisy images, as well as for restoring an under-samples images. If $x_n = x_0 + e$ is noisy observation of $x_0$, we can restore $x_0$ by following equation: min TV($x$) s.t. $\|x_n - x\|_2^2 < \varepsilon^2$, where $\varepsilon = \|e\|_2^2$. The total variation of x is sum of the gradient magnitudes at each point:

$$TV(x)=\sum_{i,j}\|D_{i,j}x\|_2 \qquad (1)$$
$$D_{i,j}x = \begin{cases} x(i+1,j)-x(i,j) \\ x(i,j+1)-x(i,j) \end{cases} \qquad (2)$$

The TV based denoising method is effective in removing the noisy pixels and reconstruction the missing pixels, while keeping the image details. The TV minimization is done according to the relation:

$$\min TV(x) \text{ s.t. } \|Ax-y\|_2^2 < \mathcal{E}^2. \qquad (3)$$

TV minimization algorithm provides a solution whose variations are concentrated on a small number of edges. The algorithm is applied to 64 x 64 blocks. The total number of samples per block is N = 4096, while the random M = 1500 measurements are used for reconstruction (the DFT matrix is used). The l1-magic toolbox is used for solving the minimization problems.

**Gradient base algorithm**

Gradient-based algorithms [12], [17] are popular when solving unconstrained optimization problems. By exploiting knowledge of the gradient of the objective function to optimize, each iteration of a gradient-based algorithm aims at approaching the minimizer of said function. In the age of web-scale prediction problems, many venerable algorithms may encounter difficulties. However, as the data sets grow larger, so does our capability to interpret them. Thanks to developments in the computer-science subfield called Machine Learning, it is possible to produce self-adapting optimizers able to deal with different challenging scenarios, without requiring the user to rewrite algorithms specialized for the experiment at hand.

**Mathematical interpretation of Gradient-Based algorithm**

A size of concentration for a certain vector x of transform coefficients is defined by:
$$\mu(x) = \sum_{k=0}^{N-1}|x(k)|. \qquad (5)$$
M measurements is modified by embedding zero values on the positions of missing samples:
$$y(n)\begin{cases} f(n), \text{for } n \in NN_M \\ 0, \text{for } n \in N/N_M \end{cases} \qquad (6)$$
In the i-th iteration, the values of vector y on the positions of missing samples are changed as follows:
$$y^+(n)=y_i(n) + \Delta\delta(n-n_s) \qquad (7)$$
$$y^-(n)=y_i(n) - \Delta\delta(n-n_s) \qquad (8)$$

The initial set up assumes $y_i(n)=y(n)$ for i=0 and n∈N. The missing samples positions are denoted as $n_s \in N\backslash N_M$, while δ is the Kronecker delta function. The initial value of the step is calculated as follows:
$$\Delta = {}^{max}_n\{|y(n)|\} \qquad (9)$$
The transform domain vectors corresponding to $y^+$ and $y^-$ are given by, where Ψ is a transform matrix:
$$X^+ = \Psi y^+ \qquad (10)$$
$$X^- = \Psi y^- \qquad (11)$$
Gradient vector is calculated as it follows, where η is concentration measure:
$$G^i(n_s)=\eta(x^+)-\eta(x^-)/N, n_s \in N\backslash N_M \qquad (12)$$
The measurement vector is then updated as:
$$y^{i+1}(n)=y^i(n)-G^i(n) \qquad (13)$$
The angle between two gradient vectors can be calculated:
$$\beta = arccos \frac{\langle G^{i-1}G^i \rangle}{\|G^{i-1}\|_2^2\|G^i\|_2^2} \qquad (14)$$
The ratio between the reconstruction error and the signal in the current iteration as follows:
$$\Xi = 10\log_{10}\frac{\sum_{n\in N\backslash N_M}|y^p(n)-y^i(n)|^2}{\sum_{n\in N\backslash N_M}|y^i(n)|^2} \qquad (15)$$

**Expirimental result**

For the given experiment, we used a bmp image, with the resolution 200x200, for three different algorithms. The given algorithms are: OMP algorithm, BP algorithm and Total-Variaton method. All three algorithms have been previously explained in the previous selection mark both in theoretical form and also in mathematical form. In the next for image we can see our result:

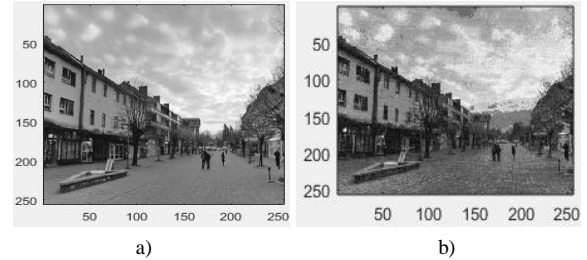

a)  b)

Figure 2. a) Original image (Berane.bmp), b) Reconstruction with BP (PSNR=30.8509dB; 90% available pixels)

The results that we get through the given algorithm for different values are shown in the following table. Performance comparison and quality estimation with PSNR show that in both cases BP algorithm is better than OMP, but the third algorithm, TV, give the best result for all three cases.

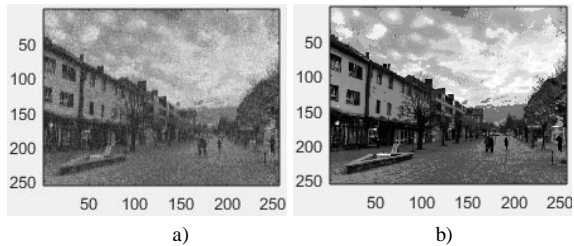

Figure 3. a) Reconstruction with OMP (PSNR=23.3835 dB), b) Reconstruction with TV (PSNR=48,49dB; 90% available pixels)

| PSNR [dB] | | | |
| --- | --- | --- | --- |
| θ | BP | OMP | TV |
| 0.1 | 10.2685 | 9.4935 | 25.63 |
| 0.3 | 19.523 | 17.8998 | 30.32 |
| 0.5 | 21.8924 | 19.9717 | 34.96 |
| 0.7 | 26.0491 | 21.7541 | 40.12 |
| 0.9 | 30.8509 | 23.3835 | 48.49 |

Figure 4. Performance comparison and quality estimation with PSNR (θ is percentage of available pixels)

**Conclusion**

A review of the compressed sensing is discussed, which demonstrates the correct choice of scanty portrayal of the signals, choice of measurement matrix for acquisition, algorithms for reconstruction and furthermore discussed compressive detecting applications. Here we can see that, algorithms utilized for the signal reconstruction center to find the arrangement of an underdetermined arrangementof linear equations using sparseness constraints. As contrasted with the conventional Nyquist-Shannon hypothesis, CS afford grind quality to the signals without expanding the amount of required information , which implies that the first signal can be reproduced with few tests (fs <2fm). This paper comprehensively describes the study of different algorithms for reproducing compressed sampled signals and compares their nature. In future work, testing of different algorithms for reconstruction of compressed sensed images would be possible.